 \documentclass[preprint2]{aastex}

\slugcomment{Accepted for publication in ApJ}

\shorttitle{Mass Ratio from H$_{\alpha}$ in Black-Hole transients}
\shortauthors{Casares}

\begin{document}

\title{Mass ratio determination from H$_{\alpha}$ lines in Black-Hole X-ray transients}

\author{J. Casares}
\affil{Instituto de Astrof\'\i{}sica de Canarias, 38205 La Laguna, 
S/C de Tenerife, Spain}
\affil{Departamento de Astrof\'\i{}sica, Universidad de La Laguna,
E-38206 La Laguna, S/C de Tenerife, Spain}
\affil{Department of Physics, Astrophysics, University of Oxford,
Keble Road, Oxford OX1 3RH, UK}

\begin{abstract}
We find that the mass ratio $q$ in quiescent black hole (BH) X-ray transients is 
tightly correlated with the ratio of the double peak separation ($DP$) to the full-width-half maximum 
($FWHM$) of the H$_{\alpha}$ emission line, $\log q = -6.88~ -23.2~\log (DP/FWHM)$. 
The correlation is explained through the efficient truncation of the outer disc radius by the 3:1 
resonance with the companion star. This is the dominant tidal interaction for extreme mass ratios 
$q=M_2/M_1\lesssim0.25$, the realm of BH (and some neutron star) X-ray transients. 
Mass ratios can thus be estimated with a typical uncertainty of $\approx$32\%, provided that the 
H$_{\alpha}$ profile used to measure $DP/FWHM$ is an 
orbital phase average.
We apply the $DP/FWHM-q$ relation to the 
three faint BH transients XTE J1650-500, XTE J1859+226 and Swift J1357-0933 and  
predict $q=0.026^{+0.038}_{-0.007}$, 0.049$^{+0.023}_{-0.012}$ and 0.040$^{+0.003}_{-0.005}$, respectively. 
This new relation, together with the $FWHM-K_2$ correlation presented in Paper I 
\citep{casares15} allows the extraction of fundamental parameters from very faint targets 
and, therefore, the extension of dynamical BH studies to much 
deeper limits than was previously possible. 
As an example, we combine our mass ratio determination for Swift J1357-0933 with previous results 
to yield a BH mass of 12.4$\pm$3.6 M$_{\odot}$. This confirms Swift J1357-0933 as one of the most massive 
BH low-mass X-ray binaries in the Galaxy.
\end{abstract}


\section{Introduction}

Stellar-mass black holes (BHs) are mostly detected through dramatic X-ray outbursts 
exhibited by transient X-ray binaries (SXTs; \citealt{tanaka96}). About 60 BH candidates have been identified 
in the 50 year lifetime of X-ray astronomy \citep{corral15}, although only 17 of these have been  
confirmed by dynamical studies i.e. they possess a mass function $f(M)=M_{1} \sin ^{3} i / (1+q)^2$ 
in excess of 3 M$_{\odot}$, where $q=M_{2}/M_{1}$ is the mass ratio of the companion star 
to the compact object. 
The reason for the low rate of confirmed BHs lies in the difficulty of detecting  
the companion star for quiescent optical magnitudes fainter than $\sim$22. 
Time-resolved spectroscopy with signal-to-noise ratio $S/N\gtrsim10$ is typically needed to  
detect the weak absorption features and trace their orbital motion. Determining the binary mass ratio 
is even more challenging as it also requires resolving powers $R\gtrsim5000$ to measure the 
rotational broadening of the absorption lines (e.g. \citealt{casares94}). In 
addition, moderately short integration times are essential to avoid significant orbital smearing. 
It should be mentioned that alternative methods based on the radial velocities of the disc emission 
lines are uncertain and prone to large systematic effects (see e.g. \citealt{marsh94}). 
A critical review on the determination of system parameters in BH SXTs can be found in \cite{casares14}. 

In \cite{casares15} (henceforth Paper I) we showed that the full-width-half maximum ($FWHM$) of the  
disc H$_{\alpha}$ line in quiescent BH SXTs scales with the velocity 
semi-amplitude of the companion star. We here present the discovery of a another correlation 
between $q$ and the ratio of the double peak separation to the line width. 
Both relations open the door to constrain fundamental parameters and perform dynamical studies 
in much fainter samples of quiescent BH candidates than is currently possible.

\section{Database}

We have selected a sample of quiescent spectra of nine dynamical BH and two neutron star (NS) SXTs from Paper I, 
all with reliable mass ratio determinations. Only mass ratios obtained by resolving the rotational broadening 
of the companion star (the $V \sin i$ technique) have been considered. Table 1 presents our 
sample and the associated references.  Full observational details for every target are given in Paper I.  
For the case of XTE J1118+480, we decided to include also 34 additional spectra obtained with the 10.4m 
Gran Telescopio Canarias (GTC) and reported in \cite{gonzalez14}. 
Note that GRO J0422+320 is not included despite having a $q$ determination through the $V \sin i$ technique. 
This is because the resolution of the only three quiescent spectra available to us is too poor ($\Delta\lambda>25$\% of the double peak 
separation) for a realistic determination of the double peak separation.  

For comparison we have also chosen a sub-sample of quiescent Cataclysmic Variables
(CVs) from Paper I. From the 24 possible CVs with reliable $q$ determinations, 13 have 
been rejected because the line core is dominated by a strong S-wave component which 
prevents us from measuring the double peak separation with the simple model outlined below. 
In any case, these are all CVs with large mass ratios $q\gtrsim0.6$ which, as we will show later, are not 
relevant to our analysis.  
From the remaining 11 CVs, three have $q$ values based on the  $V \sin i$ technique (GK Per, IP Peg and CTCV J1300-3052) 
and another two on a direct determination of the radial velocity curves of both the white dwarf and the donor star (U Gem 
and WZ Sge). In the remaining six the  companion star  is not directly observed but robust $q$ values 
are available through modeling the eclipse of the white dwarf and the hot spot (see e.g. \citealt{littlefair08}).  
Our CV sample is listed in Table 2.

\section{The $DP/FWHM-q$ relation} 

Double peak separations ($DP$) were obtained by fitting a symmetric 2-Gaussian model 
to the average H$_{\alpha}$ profile in every SXT and CV. 
In the case of eclipsing CVs we excluded those spectra obtained within $\pm$0.05 phases from the time of the central eclipse.  
The fitted model consists of a constant plus two Gaussians of identical width and height. 
The continuum rectified spectra were fitted in a window of $\pm$10000 km s$^{-1}$, 
centered on the H$_{\alpha}$ 
line after masking the neighboring HeI line at $\lambda$6678. 
Prior to the fit, the 2-Gaussian model was degraded to the resolution of the data by convolution with the 
instrumental profile\footnote{As a test we also tried fitting the model without instrument degradation and find that this only 
has a minor impact ($\lesssim1$\%) on the ratio $DP/FWHM$ because both quantities are almost equally affected.} 
We adopted 1-$\sigma$ 
formal errors on the fitted parameter as derived through $\chi^2$ minimization. Fig. 1 
displays some fit examples using our 2-Gaussian model. 
In addition, $FWHM$ values were 
extracted from single Gaussian fits to the same average H$_{\alpha}$ profiles following Paper I. 

 Tables 1 and 2 list the parameter $DP/FWHM$ and its propagated 1-$\sigma$ error as 
 derived from our Gaussian model fits. The evolution of $DP/FWHM$ with $q$ is presented 
 in Fig. 2. The figure shows that $DP/FWHM$ varies very rapidly for small $q$ values, 
 with a 10\% increase for $q$ under 0.2.
 
 In order to understand this behaviour, 
 we follow on from Paper I and start by assuming that the $FWHM$ of the 
 H$_{\alpha}$ line is determined by gas with Keplerian velocity at a characteristic radius 
 $R_{\rm W}=\alpha R_{L1}$, with $\alpha<1$ and $R_{L1}$ the Roche lobe of the compact 
 star, i.e.

\begin{equation}
\frac{FWHM}{2} = \left( \frac{G M_{1}}{R_{\rm W}}\right) ^{1/2} \sin i .
\end{equation}

On the other hand, the double peak separation is set by the velocity of the outer disc, 
whose radius $R_{\rm d}$ is truncated by the tidal forces of the companion star 
\citep{paczynski77, papaloizou77}. There is ample evidence for the outer disc velocities 
to be sub-Keplerian (e.g.  \citealt{north02}) and thus we decided to adopt

\begin{equation}
\frac{DP}{2} =\beta \left( \frac{G M_{1}}{R_{\rm d}}\right)^{1/2} \sin i .
\end{equation}

\noindent 
where the parameter $\beta<1$ accounts for the fraction by which the outer disc 
material is sub-Keplerian. 
For extreme mass ratios $q\lesssim0.25$ the disc is effectively truncated at the resonance radius 
of the (j=3, k=2) commensurability or   
3:1 resonance radius i.e. the radius at which the disc angular velocity is three times the angular 
velocity of the companion star (see \citealt{hirose90, frank02}). Therefore, 

\begin{equation}
R_{\rm d} \equiv R_{32}= 3^{-2/3} \left(1+q\right)^{-1/3} a 
\end{equation}
 
\noindent
where $a$ the binary separation. If we now bring eq. 3 into eq. 2 
and use Eggleton's relation \citep{eggleton83} to remove $R_{L1}/a$ we find

\begin{equation}
\frac{DP}{FWHM}=3^{1/3} \left(1+q\right)^{2/3} \beta \sqrt{\alpha  f(q)}  
\end{equation}

\noindent
where $f(q)$ is the same expression as in eq. 6 of Paper I, i.e.

\begin{equation}
f(q)= \frac{0.49 \left(1+q \right)^{-1}}{0.6 + q^{2/3} \ln \left(1+q^{-1/3}\right)} .
\end{equation}
 
By computing the ratio between the double peak separation and the 
line width we have managed to cancel out the dependence on compact 
object mass and binary inclination. Interestingly, in contrast with the $FWHM-K_2$ 
correlation presented in Paper I,  eq. 4 is very sensitive to the mass ratio for  
$q\lesssim0.25$ i.e. the typical values achieved by BH SXTs.  

For the sake of comparison, we also plot eq. 4 in Fig. 2 for  $\alpha=0.42$ (adopted from Paper I) 
and $\beta=0.77$.  Although not intended to be a 
fit, the alignment of the data with the model indicates that eq. 4 provides a good description of the observations. 
This endorses our interpretation that the strong dependence of  $DP/FWHM$ with $q$ 
is driven by the truncation of the outer disc radius caused by the 3:1 resonance tide.  
The bottom panel in Fig. 2 also displays the evolution of $DP/FWHM$ 
for the CV sample, with the model for $\alpha=0.42$ and $\beta=0.83$ superimposed. 
The comparison of the SXT and CV data with the model indicates that the velocities at the outer 
accretion disc are sub-Keplerian by $\approx$20\%, in good agreement with 
other studies (e.g. \citealt{wade88}). Note that for $q\gtrsim0.25$ the 3:1 resonance lies 
beyond $\approx$0.9 $R_{L1}$ and the outer disc radius 
is then limited by the largest non-intersecting orbits allowed by three-body 
interactions \citep{paczynski77}. Under these circumstances, the dependence of the outer disc radius 
on $q$ is very weak \citep{frank02} and, therefore, a nearly constant evolution of 
$DP/FWHM$ versus $q$ is expected for  $q\gtrsim0.25$.

For practical purposes, we display in Fig. 3 the variation of $DP/FWHM$ versus $q$ in logarithmic 
units. A least-squares linear fit yields 

\begin{equation}
\log q= -6.88 (0.52) - 23.2 (2.0)  \log\left(\frac{DP}{FWHM}\right) 
\end{equation}

\noindent 
witha Pearson correlation coefficient $r=0.95$. In order to estimate the error in $q$ 
implied by this relation we have computed the difference with respect to the true 
observed values for our 11 SXTs. The distribution of differences can be approximated by 
a normal function with $\sigma (q)=0.015$. This indicates that mass ratios can be 
realistically obtained from eq. 6 with a typical $\sim25$\% uncertainty.

\section{Orbital effects}

The spectra that we have used to produce the $DP/FWHM-q$ correlation are orbital 
averages. This is party because individual spectra rarely possess enough signal-to-noise 
for the technique to be applicable. But also, because orbital means average out 
possible asymmetries in individual spectra from, for example, hot-spots or disc 
eccentricitiies which could potentially bias the determination of $q$. 

At this point we 
decided to explore the impact of line asymmetries in the results of our technique. 
Since this can only be tested on data with sufficient signal-to-noise we have focused 
on the 154 high quality GTC spectra of XTE J1118+480 obtained along four different orbits 
over two years. We have performed Gaussian fits on every individual spectrum and 
computed mass ratios using eq. 6. The distribution of $q$ values is found to peak at 
$q=0.024$ (bottom panel in Fig. 4), with 68\% of the values contained between $q=0.021$ and 0.045.  
This indicates that, if $q$ were to be obtained from a single individual spectrum, the 
typical uncertainty would be about $\sim$56\%. 

In order to trace the effect of line asymmetries we also extracted the $V/R$ parameter 
from each spectrum, with $V/R$ defined as the ratio of equivalent widths between the blue and the red part of 
the H$_{\alpha}$ profile. The two halves of the line are set from the rest wavelength till $\pm$2500 km s$^{-1}$. 
For example, $V/R=0.8$ indicates a line with the red part stronger by 20\% while $V/R=1$ implies 
a symmetric profile. A plot of $q$ versus $V/R$ (see top panel in Fig. 4) seems to show a trend, with  
asymmetric profiles preferring slightly lower $q$ values, although the large scatter prevents from 
drawing a firm conclusion.    

However, as we have mentioned above, the technique outlined in this paper is most useful on phase 
averaged spectra because individual spectra typically have very limited signal-to-noise. 
Consequently, we have extracted $q$ values by fitting Gaussians to the four orbital averages  
of the 154 individual spectra. The distribution of $q$ values has a mean at 0.026 and a 
standard deviation of 0.005, indicating that the typical error on phase averaged spectra is 
reduced to  $\sim$19\% i.e. smaller than the 25\% uncertainty drawn from the correlation. 
We therefore conclude that the uncertainty expected from the application of our technique 
to phase averaged spectra is about 32\%.

\section{Discussion: application to three faint BHs}

In Paper I we showed that the $FWHM$ of the H$_{\alpha}$ line in quiescent SXTs and 
CVs is formed at $\simeq$42\% of $R_{L1}$. Furthermore, it is tightly correlated with 
the projected velocity of the donor star $K_{2}$ and thus, the quantity $FWHM/K_2$ 
can be used to extract dynamical information from single epoch low resolution 
($R\gtrsim500$) spectroscopy. In addition, we showed that $FWHM/K_2$ is weakly dependent on $q$,  
resulting in a $\sim$27\% flatter slope for (long-period) CVs.  

We here now present a new method to estimate the binary mass ratio in quiescent BH SXTs from the 
properties of the H$_{\alpha}$ line.  We have proved that the quantity $DP/FWHM$ is strongly dependent on $q$, 
with a 10\% variation for $q\lesssim0.25$. The reason behind this is the efficient truncation of the  
outer disc radius by the 3:1 tidal resonance of the donor star. The correlation of  $DP/FWHM$ with $q$, 
therefore, opens a new avenue to measure mass ratios in quiescent BH SXTs. The double peak 
separation can be solidly measured by fitting a symmetric double-Gaussian model to 
phase averaged  H$_{\alpha}$ profiles. We estimate that instrumental resolution better than 25\% of the double peak 
 separation is required to resolve the latter. This typically demands resolving powers of only 
$R\gtrsim$1000 i.e a factor $\sim5$ lower than required to measure $q$ using the $V \sin i$ technique. 
More significantly for observational feasibility, this method makes use of the disc H$_{\alpha}$ line which, with a 
typical $EW\sim50$~\AA, is much stronger that the weak atmospheric features of the donor star.   

The relations presented in this paper and in Paper I thus allow for a reasonably accurate estimation of the system parameters in very faint 
SXTs which otherwise cannot be tackled with current instrumentation and standard techniques. 
As an example, we have applied our method to the BH SXTs XTE J1650-500, XTE J1859+226 and 
Swift J1357-0933. They all have $R\simeq22-23$ and none of them has yet a mass ratio determination.  
We have produced averaged spectra for XTE J1650-500 and XTE J1859+226 using the data 
presented in Table 1 of Paper I. Regarding Swift J1357-0933 we have used a more recent and extended database 
reported in \cite{mata15}. Fig.5 displays the averaged spectra of the three BHs together with the best double-Gaussian model 
fits, which result in $DP/FWHM=0.5828\pm0.0150$, 0.5741$\pm$0.0072 and 0.5805$\pm$0.0027 for XTE J1650-500, 
XTE J1859+226 and Swift J1357-0933 respectively. The mass ratios implied by eq. 6 are 
$q=0.026^{+0.038}_{-0.007}$, 0.049$^{+0.023}_{-0.012}$ and 0.040$^{+0.003}_{-0.005}$, respectively. 
The quoted uncertainties correspond to 68\% confidence regions and have been computed using a Monte Carlo simulation 
with $10^5$ realizations. 
We note in passing that our mass ratio for Swift J1357-0933 is in excellent agreement with an independent estimate 
based on the radial velocity curve of the wings of the H$_{\alpha}$ line \citep{mata15}. 

Regarding Swift J1357-0933 we are now in a position to present a credible BH mass based on our 
scaling relations. By combining our mass ratio with the mass function obtained by means of the $FWHM/K_2$ correlation 
\citep{mata15} and a conservative estimate of the inclination angle $i=80\pm10^{\circ}$ \citep{corral13,mata15} we find 
$M_1=12.4\pm3.6$ M$_{\odot}$. This result confirms Swift J1357-0933 as one of the most massive BH low-mass 
X-ray binaries in the Galaxy, only rivaled by GRS 1915+105 \citep{reid14}.

To conclude we would like to stress that the relations presented here and in Paper I will 
help deepen the search for new BH transients to substantially fainter limits. They will 
also prove very useful in extracting fundamental parameters from large numbers of 
data to be delivered by 
new spectroscopic surveys such as GAIA or WEAVE. 

\acknowledgments
We would like to acknowledge the hospitality of the Department of Physics of the 
University of Oxford, where this work was performed during a sabbatical visit. 
We also thank the anonymous referee for valuable comments which 
helped to improve the quality of the paper. 
We are grateful to Serena Repetto for bringing to our attention a mass ratio determination 
based on the $V \sin i$ technique for N. Vel 93 (=GRS 1009-45). 
We further thank P. Charles, T. Maccarone and T. Mu\~noz-Darias for useful 
comments on the manuscript of this paper an on that of Paper I. 
This paper makes use of data obtained from the Isaac Newton Group Archive 
which is maintained as part of the CASU Astronomical Data Centre at the 
Institute of Astronomy, Cambridge.
Partially based on observations made with the GTC operated on the island of La Palma 
by the Instituto de Astrof\'\i{}sica de Canarias in the Spanish Observatorio del Roque de Los Muchachos
of the Instituto de Astrof\'\i{}sica de Canarias.
This work is supported by DGI of the Spanish Ministerio de Educaci\'on, Cultura y 
Deporte under grants AYA2013-42627 and PR2015-00397.

\clearpage

\begin{figure}
\includegraphics[angle=-90,scale=0.55]{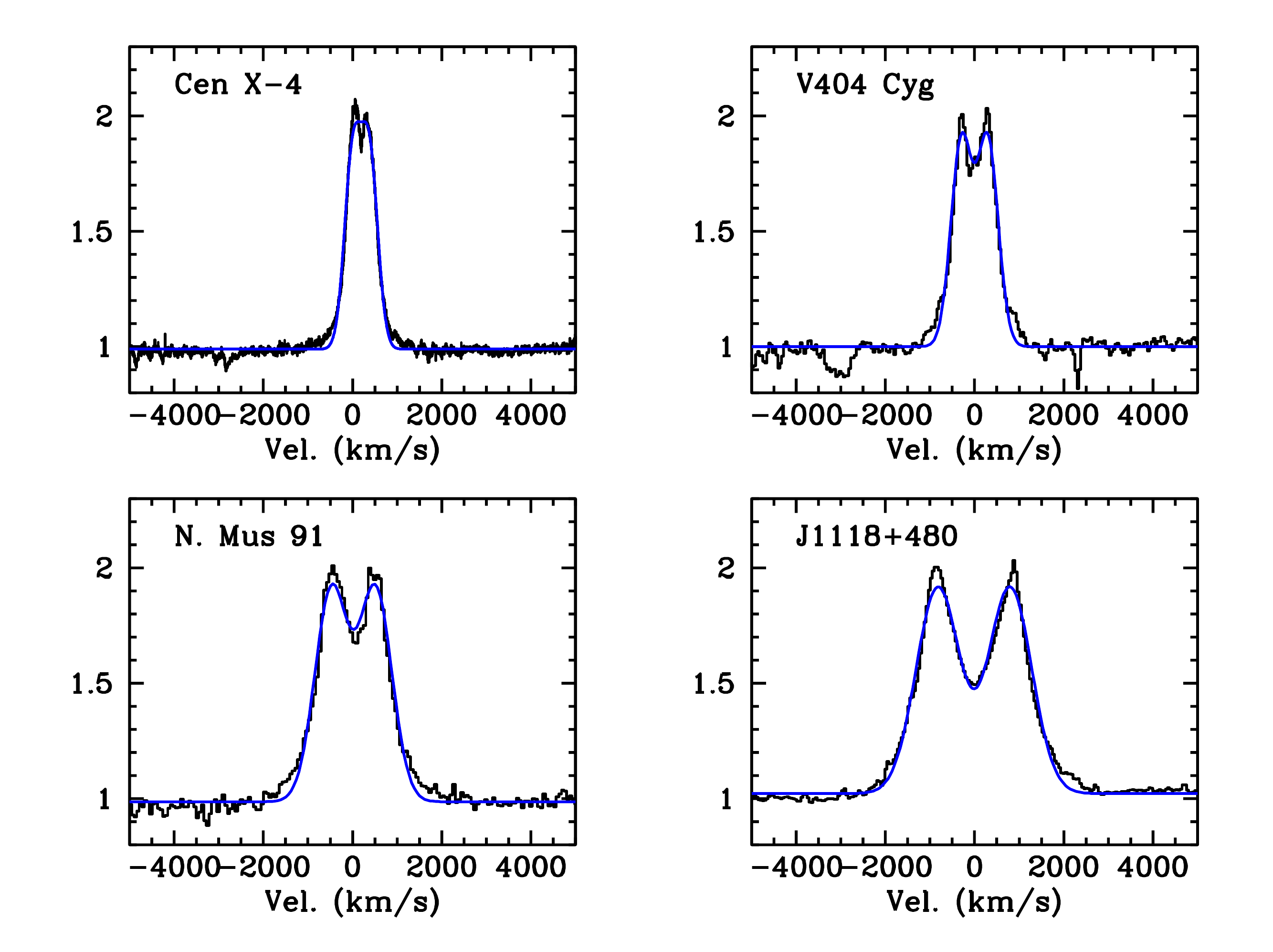}
\caption{Example of double Gaussian fits to H$\alpha$ profiles in SXTs. A selection of average spectra, representing the entire range of  
$FWHM$s, is depicted. 
\label{fig1}}
\end{figure}

\clearpage

\begin{figure}
\includegraphics[angle=0,scale=0.55]{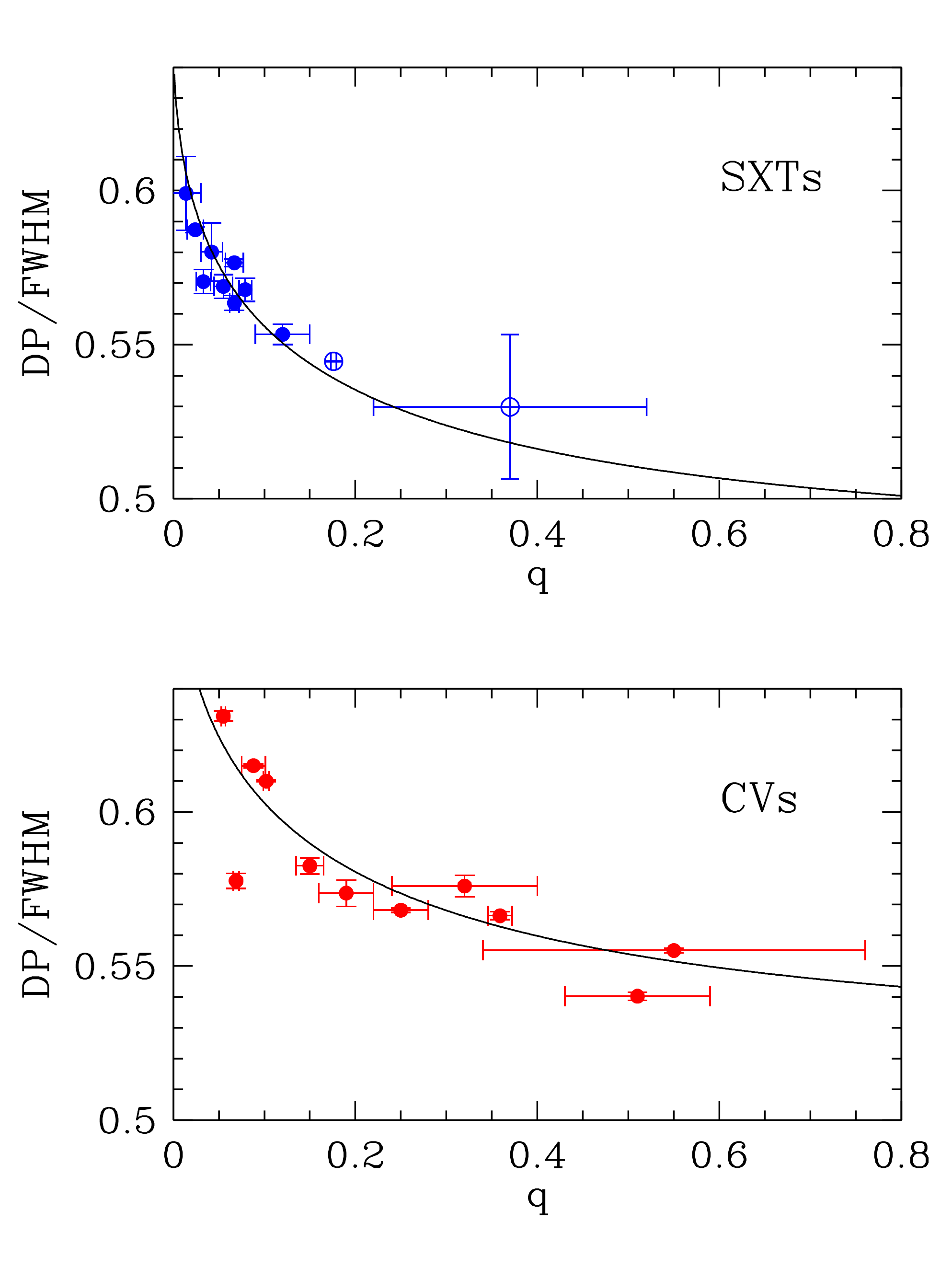}
\caption{Top: evolution of the  $DP/FWHM$ parameter with mass ratio $q$ for SXTs. 
Blue solid circles indicate BHs while NSs are marked by open circles. The solid line represents 
eq. 4 for $\alpha=0.42$ and $\beta=0.77$. 
Bottom: same as above for  CVs. The solid line depicts 
eq. 4 for $\alpha=0.42$ and $\beta=0.83$. 
\label{fig2}}
\end{figure}

\clearpage

\begin{figure}
\includegraphics[angle=-90,scale=.60]{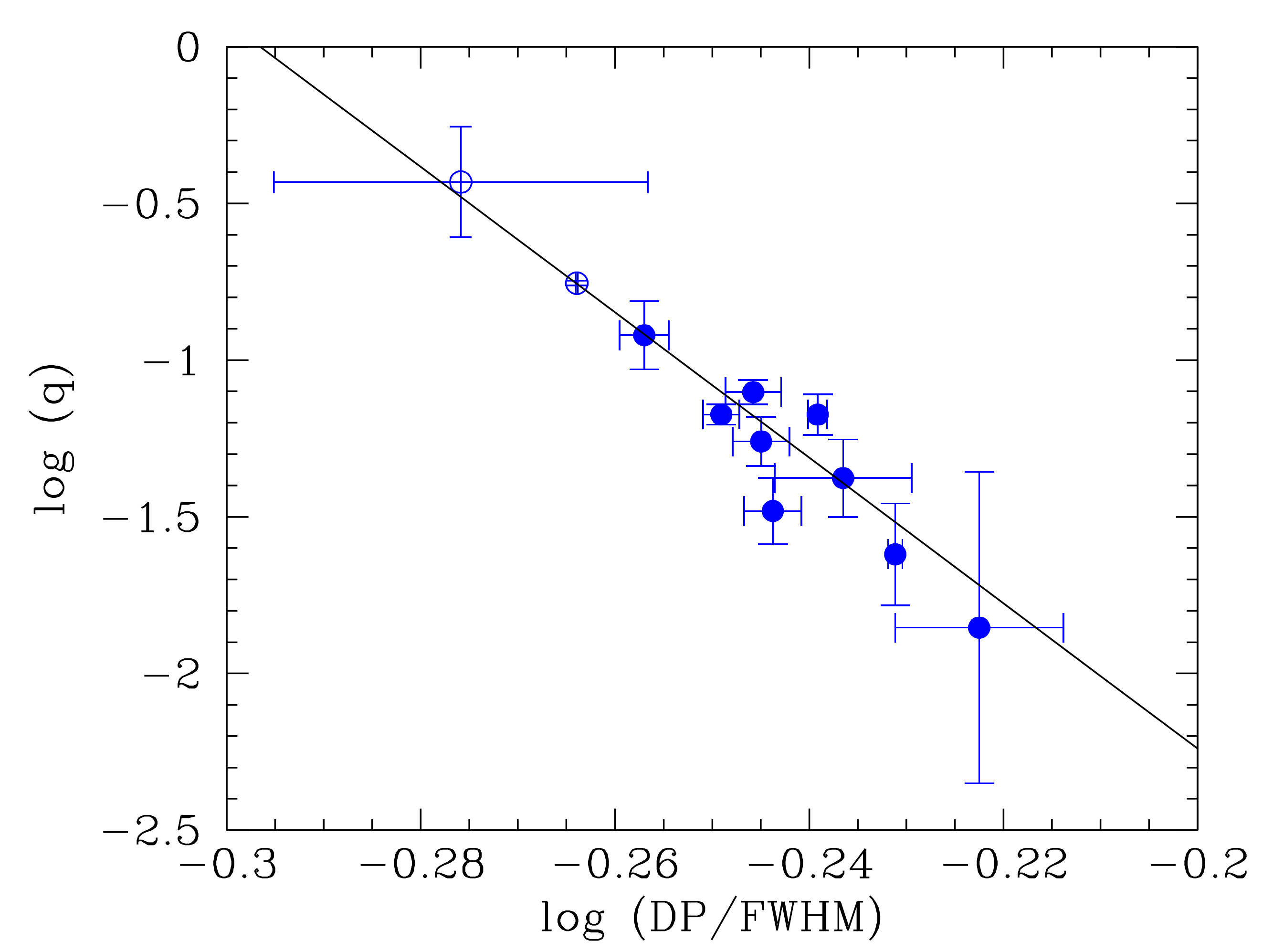}
\caption{The correlation between $q$ and the parameter $DP/FWHM$ in log units and the best linear fit. 
\label{fig3}}
\end{figure}

\clearpage

\begin{figure}
\includegraphics[angle=0,scale=.60]{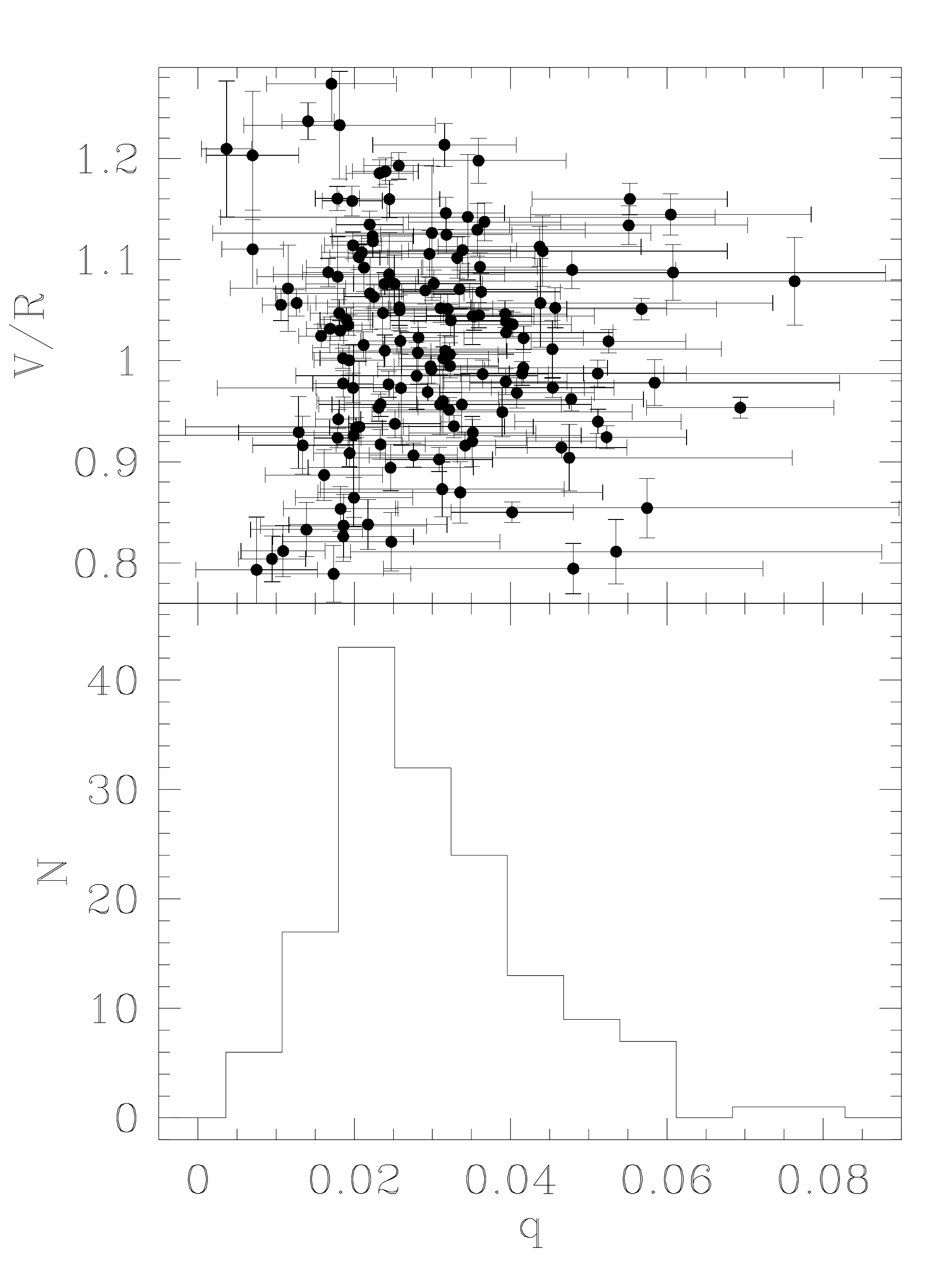}
\caption{Bottom: histogram of $q$ values derived through eq.6 from 154 GTC spectra of XTE J1118+480. Top: variation $q$ with the $V/R$ parameter of 
the H$\alpha$ profile (see text for details).
\label{fig4}}
\end{figure}

\clearpage

\begin{figure}
\includegraphics[angle=0,scale=.60]{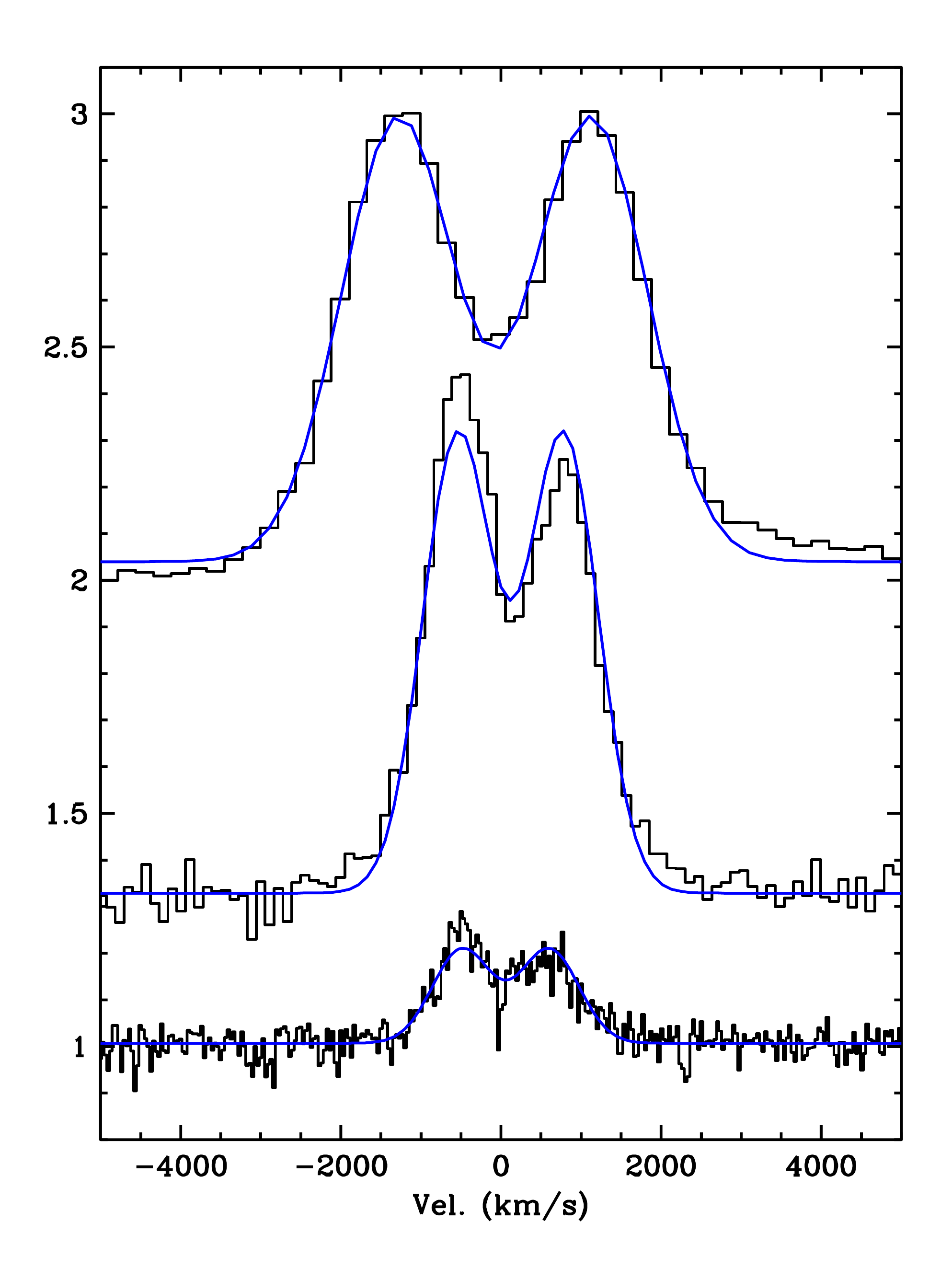}
\caption{From bottom to top: average H$\alpha$ profiles of XTE J1650-500, XTE J1859+226 and Swift J1357-0933 with the best two-Gaussian model fits.
\label{fig5}}
\end{figure}

\clearpage

\begin{deluxetable}{lccc}
\tablewidth{0pt}
\tablecaption{Database of X-ray transients\label{tbl-1}}
\tablehead{
\colhead{Object} & \colhead{q} & \colhead{DP/FWHM}  & \colhead{ref} } 
\startdata
\cutinhead{Black Holes}
V404 Cyg             &  0.067 $\pm$  0.005  & 0.5636 $\pm$  0.0024 & 1 \\
BW Cir                 &  0.12   $\pm$  0.03    & 0.5534  $\pm$ 0.0032 & 2 \\
XTE J1550 -564  &  0.033  $\pm$ 0.008  & 0.5705  $\pm$ 0.0039 & 3 \\
N. Oph 77            &  0.014 $\pm$  0.016  & 0.5991  $\pm$ 0.0119 & 4 \\
N. Mus 91            &  0.079  $\pm$ 0.007    & 0.5679  $\pm$ 0.0037 & 5 \\
GS 2000+25        &  0.042 $\pm$  0.012  & 0.5801  $\pm$ 0.0094 & 6 \\ 
A0620-00             &  0.067 $\pm$  0.010  & 0.5766  $\pm$ 0.0013 & 7 \\ 
N Vel 93               & 0.055  $\pm$ 0.010   & 0.5689  $\pm$ 0.0038 & 8 \\
XTE J1118+480    & 0.024  $\pm$  0.009  & 0.5873 $\pm$ 0.0010 & 9 \\ 
\cutinhead{Neutron Stars}
Cen X-4               &  0.176  $\pm$  0.003  & 0.5446 $\pm$  0.0001 & 10 \\ 
XTE J2123-058   &  0.37    $\pm$  0.15    & 0.5298 $\pm$  0.0234 & 11 \\ 
\enddata
\tablerefs{
(1) \cite{casares96}; (2) \cite{casares09a}; (3) \cite{orosz11}; 
(4) \cite{harlaftis97}; (5) \cite{wu15}; (6) \cite{harlaftis96}; (7) 
\cite{marsh94}; (8) \cite{macias11}; (9) \cite{calvelo09}; (10) \cite{shahbaz14}; 
(11) \cite{tomsick02}
.}
\end{deluxetable}

\clearpage

\begin{deluxetable}{lccc}
\tablewidth{0pt}
\tablecaption{Database of Cataclysmic Variables.\label{tbl-2}}
\tablehead{
\colhead{Object} & \colhead{q} & \colhead{DP/FWHM}  & \colhead{ref} } 
\startdata
GK Per                                   &  0.55    $\pm$  0.21    & 0.5551  $\pm$ 0.0007 &  1 \\ 
SDSS J100658.40+233724   &  0.51     $\pm$  0.08    & 0.5402   $\pm$  0.0013 & 2 \\
U Gem                                   &  0.359   $\pm$  0.013  & 0.5663   $\pm$  0.0013  & 3,4 \\
IP Peg                                    &  0.32     $\pm$  0.08    & 0.5760  $\pm$   0.0035  & 5 \\
CTCV J1300-3052                 &  0.25     $\pm$  0.03    & 0.5681  $\pm$   0.0007 & 6 \\
HT Cas                                   &  0.150   $\pm$  0.015  & 0.5825   $\pm$  0.0027 & 7 \\
OY Car                                   &  0.102   $\pm$  0.003  & 0.6100   $\pm$  0.0003  & 8\\
V2051 Oph                             &  0.19    $\pm$   0.03    & 0.5737   $\pm$  0.0043 & 9 \\
SDSS 103533.02+055158.3  &  0.055  $\pm$   0.002  & 0.6311   $\pm$  0.0016  & 8 \\
WZ Sge                                  &  0.088  $\pm$   0.013  & 0.6150   $\pm$  0.0007 &  10 \\
SDSS J143317.78+101123.3 &  0.069  $\pm$   0.003  & 0.5777  $\pm$   0.0025 & 8 \\
\enddata
\tablecomments{$q$ values for GK Per, IP Peg and CTCV J1300-3052 have been obtained through the $V \sin i$ technique while 
those for U Gem and WZ Sge by measuring the radial velocity curves of the white dwarf and the donor star. The 
remaining $q$ values are derived by modeling the eclipses of the white dwarf and the hot-spot in optical light curves. 
}
\tablerefs{
(1) \cite{morales-rueda02}; (2) \cite{southworth09}; 
(3) \cite{friend90}; (4)  \cite{long99}; (5) \cite{beekman00}; 
(6)  \cite{savoury12}; (7) \cite{wood90}; (8) \cite{littlefair08}; 
(9) \cite{baptista98}; (10) \cite{steeghs07}
 .}
\end{deluxetable}

\end{document}